\documentclass[lettersize,journal]{IEEEtran}
\usepackage{amsmath,amsfonts}
\usepackage{algcompatible}
\usepackage{algorithm}
\usepackage{array}
\usepackage[caption=false,font=normalsize,labelfont=sf,textfont=sf]{subfig}
\usepackage{textcomp}
\usepackage{stfloats}
\usepackage{url}
\usepackage{verbatim}
\usepackage{graphicx}
\usepackage{multirow}
\usepackage[table]{xcolor}
\usepackage{cite}
\usepackage{pifont}
\usepackage{amssymb}
\usepackage{algpseudocode}
\usepackage{booktabs}
\usepackage{multirow}
\usepackage{makecell}

\begin{document}
\title{LoRA-Key: User-Centric LoRA Watermarking for Text-to-Image Diffusion Models}

\author{
Yaopeng Wang, Qingliang Wang, Zhibo Wang,~\IEEEmembership{Senior Member,~IEEE}, Huiyu Xu, Jiacheng Du, \\Qiu Wang, Jia-Li Yin,~\IEEEmembership{Member,~IEEE}, and Kui Ren,~\IEEEmembership{Fellow,~IEEE}

\thanks{Yaopeng Wang is with the School of Cyber Science and Engineering, Southeast University, Nanjing 211189, China, and also with the State Key Laboratory of Blockchain and Data Security, Zhejiang University, Hangzhou 310027, China(e-mail: yaopengwang@seu.edu.cn).

Huiyu Xu, Zhibo Wang, Jiacheng Du, and Qiu Wang are with the State Key Laboratory of Blockchain and Data Security, Zhejiang University, Hangzhou 310027, China, and also with the Hangzhou High-Tech Zone (Binjiang) Institute of Blockchain and Data Security, Hangzhou, Zhejiang 310012, China (e-mail: \{huiyuxu, zhibowang, jcdu, wang2qiu\}@zju.edu.cn).

Qingliang Wang and Jia-Li Yin are with the College of Computer and Data Science, Fuzhou University, Fuzhou, Fujian 350108, China (e-mail: yanjiu8888@gmail.com, jlyin@fzu.edu.cn).

Kui Ren is with the State Key Laboratory of Blockchain and Data Security, Zhejiang University, Hangzhou 310027, China, and also with the School of Cyber Science and Engineering, Southeast University, Nanjing 211189, China (e-mail: kuiren@zju.edu.cn).}
\thanks{
Corresponding author: Kui Ren.
} 
}



\maketitle

\begin{abstract}
Low-Rank Adaptation~(LoRA) has been widely adopted for customizing text-to-image diffusion models, enabling low-cost adaptation through lightweight and portable modules that are increasingly shared, reused, and commercialized on platforms such as Civitai and HuggingFace. 
This emerging LoRA assets ecosystem shifts the practical focus of copyright protection from large-scale foundation models to independently distributed LoRA modules, making LoRA copyright protection practical and urgent.
However, existing methods either protect the foundation diffusion model, leaving LoRA assets vulnerable to unauthorized copying and redistribution, or depend on watermark retraining tailored to each specific LoRA, making them impractical for open community settings where diverse creators continuously publish new LoRA assets.
To address this limitation, we propose \textbf{LoRA-Key}, a user-centric LoRA watermarking method that treats copyright protection as a reusable ownership key instead of retraining a new watermark for each LoRA asset.
LoRA-Key protects customized LoRA assets by attaching a user-specific Watermark LoRA, which carries a recoverable secret message for identifying the creator. Once trained, the Watermark LoRA can be combined with different target LoRAs through training-free linear superposition, without per-LoRA retraining or structural modification. To train such a reusable key, we first establish a latent watermark prior in the frozen VAE latent space for robust message embedding and recovery, and then optimize the Watermark LoRA with message-conditioned watermark supervision and semantic consistency constraints. We further introduce Gradient Orthogonal Projection (GOP) to suppress watermark updates that conflict with semantic-preserving directions, thereby reducing interference with generation fidelity and downstream style adaptation. 
Extensive experiments demonstrate that \textbf{LoRA-Key} provides lightweight plug-and-play copyright protection while preserving generation quality and style fidelity, and maintains robust ownership verification under image-level distortions, downstream fine-tuning, and multi-LoRA composition.

\end{abstract}

\begin{IEEEkeywords}
Diffusion watermarking, Low-Rank Adaptation, Copyright protection, Text-to-image generation, Model ownership verification

\end{IEEEkeywords}



\section{Introduction}
Low-Rank Adaptation (LoRA)~\cite{hu2022lora} has become a key mechanism for customizing text-to-image (T2I) diffusion models, enabling users to adapt powerful foundation models to specific styles and personalized generation tasks without full-parameter fine-tuning~\cite{han2024parameter, shenaj2026not, guo2024intlora, ruiz2023dreambooth, kumari2023multi, gal2022image}. 
Beyond its technical efficiency, LoRA has also changed how users share and commercialize customized diffusion models: lightweight LoRA modules can now be independently released, reused, and monetized on community-driven platforms such as Civitai and HuggingFace~\cite{wu2024mixture, civitai, huggingface_lora, huggingface2020}. 
As lightweight LoRA modules are increasingly released and reused as independent assets, the target of copyright protection is shifting from the foundation models to LoRA modules that creators actually share and commercialize~\cite{li2025rethinking, yang2025promptcos}.
This shift introduces a new copyright protection challenge: LoRA modules are easier to copy and reuse without authorization, because their lightweight and modular nature makes them highly transferable.
Although watermarking has been extensively studied for diffusion models, most existing methods are designed to protect generated images, foundation models, or the full generation pipeline~\cite{blattmann2023stable, wen2023tree, feng2024aqualora, wang2025sleepermark}. Such protection targets are increasingly mismatched with the LoRA-centric deployment paradigm. 
In this setting, an adversary can bypass foundation-model protection by extracting or reusing the LoRA module and attaching it to a clean, unprotected base model.

Recent studies have begun to explore watermarking mechanisms specifically designed for LoRA modules. 
For example, AuthenLoRA~\cite{shi2025authenlora} incorporates watermark embedding directly into stylized LoRA training, jointly optimizing stylization quality and watermark propagation to generated images. 
We refer to this paradigm as task-centric LoRA watermarking, where the watermark is learned together with a specific style adaptation task and must be optimized separately for each target LoRA. 
This per-LoRA training paradigm fundamentally limits scalability and deployment flexibility in real-world LoRA sharing settings, where creators may continuously release new LoRA assets. 
Moreover, such task-centric protection still ties the ownership signal to each specific LoRA training process, rather than providing a reusable ownership mechanism. 
This coupling can interfere with the intended stylization behavior and weaken robustness under LoRA composition or downstream adaptation~\cite{gan2023towards}.


To address these issues, we propose \textbf{LoRA-Key}, a user-centric LoRA watermarking framework that treats copyright protection as a reusable user-side ownership key rather than a task-specific watermark. Instead of embedding the watermark into each target style LoRA through repeated retraining, LoRA-Key encapsulates ownership signals into a dedicated Watermark LoRA that can be attached to independently trained LoRA assets in a plug-and-play manner. Specifically, we first establish a latent watermark prior for message embedding and extraction using the frozen VAE of the base diffusion model. We then optimize the Watermark LoRA under this message-conditioned prior together with a semantic consistency objective, where a Gradient Orthogonal Projection (GOP) mechanism explicitly removes the first-order component of watermark updates that conflicts with semantic-preserving directions. As a result, the learned Watermark LoRA becomes substantially less entangled with downstream style adaptation and can be directly combined with user LoRAs through training-free linear superposition, without requiring per-LoRA watermark retraining. Finally, ownership is verified by recovering the secret message from a suspicious image. In this way, LoRA-Key turns copyright protection into a reusable and composable module that better matches the practical distribution, reuse, and composition patterns of LoRA assets.

The main contributions of this paper are summarized as follows:
\begin{itemize}

    \item We propose a user-centric LoRA copyright protection paradigm, where a single reusable Watermark LoRA can protect multiple task-specific LoRA assets for one user.

    \item We propose LoRA-Key, a plug-and-play lightweight watermarking framework that uses Gradient Orthogonal Projection to disentangle the watermark pattern from style semantics, enabling its injection into arbitrary LoRAs without retraining.

    \item Extensive experiments demonstrate that LoRA-Key preserves generation quality and style fidelity while enabling robust ownership verification under image distortions, downstream fine-tuning, and multi-LoRA composition.
\end{itemize}




\section{Related Work and Background}

\subsection{Latent Diffusion Models and LoRA Adaptation}
Latent Diffusion Models (LDMs) perform text-to-image generation in a compressed latent space, where a frozen variational autoencoder maps images into latent representations and a denoising network iteratively predicts noise conditioned on text prompts~\cite{rombach2022high}. Given an initial noise sample $z_T$ and a text condition $c$, the generation process can be written as
\begin{equation}
    I = G(z_T, c; \Theta),
\end{equation}
where $\Theta$ denotes the parameters of the deployed diffusion model.

Low-Rank Adaptation (LoRA)~\cite{hu2022lora} has become a widely used parameter-efficient mechanism for customizing diffusion models. Instead of updating the full weight matrix $W_0 \in \mathbb{R}^{d \times k}$, LoRA freezes the original parameters and represents the task-specific update as a low-rank decomposition:
\begin{equation}
    \Delta W = BA,
\end{equation}
where $B \in \mathbb{R}^{d \times r}$ and $A \in \mathbb{R}^{r \times k}$, with $r \ll \min(d,k)$. The modified forward pass is therefore given by
\begin{equation}
    h_{\mathrm{out}} = (W_0 + \Delta W) h_{\mathrm{in}} = (W_0 + BA) h_{\mathrm{in}}.
\end{equation}
This lightweight parameterization enables users to efficiently adapt a pre-trained diffusion model to specific subjects, concepts, or artistic styles.

In practical deployment, LoRA modules are usually treated as independent assets that can be shared, reused, and composed with a frozen base model. Let $\Theta_0$ denote the base diffusion model and $\Delta \Theta_s$ denote a user-trained style LoRA. The deployed model can be formulated as
\begin{equation}
    \Theta_{\mathrm{style}} = \Theta_0 + \alpha \Delta \Theta_s,
\end{equation}
where $\alpha$ controls the strength of the style LoRA. More generally, multiple LoRA modules can be combined through arithmetic superposition:
\begin{equation}
    \Theta_{\mathrm{merge}} = \Theta_0 + \sum_{i=1}^{N} \alpha_i \Delta \Theta_i .
\end{equation}

Such compositional deployment is common in LoRA-based image generation, where users combine independently trained adapters to obtain richer semantic or stylistic control~\cite{huang2023lorahub,gu2023mix,wu2024mixture}. However, LoRA composition is not always benign~\cite{ilharco2022editing,yadav2023ties}. Since independently trained low-rank updates may occupy overlapping or conflicting directions in parameter space, naive weight addition can introduce destructive interference and degrade generation quality~\cite{shah2024ziplora,shenaj2025lora}. This issue is particularly important for LoRA watermarking: a watermark module must remain detectable after being composed with arbitrary target LoRAs, while also avoiding interference with the original style or subject adaptation. Therefore, copyright protection for LoRA assets requires not only robust watermark embedding, but also compatibility with modular LoRA composition.

\subsection{Watermarking Methods for Diffusion Models}

Watermarking has been widely studied for identifying the provenance of generated images and verifying the ownership of generative models~\cite{shao2026reading,shao2024explanation,li2025move}. Existing diffusion watermarking methods can be broadly divided into image-level and model-level approaches. Image-level methods embed ownership signals directly into generated images, often through spatial-domain, frequency-domain, or neural watermark encoders~\cite{xia1998wavelet,zhu2018hidden,cox2007digital,tancik2020stegastamp,zhang2019robust}. Model-level methods inject watermark signals into the diffusion model, the decoder, or the generation trajectory, so that generated images implicitly carry recoverable ownership information~\cite{fernandez2023stable, wen2023tree,wang2025pt,yang2024gaussian,huang2024robin,feng2024aqualora,zhao2023recipe,zhang2024attack}.

Although these methods have demonstrated promising robustness against image-space distortions or model-level manipulations, most of them are designed under a monolithic deployment assumption, where the protected object is the foundation model, the decoder, or the complete generation pipeline. This assumption becomes restrictive in the emerging LoRA asset ecosystem. In practice, high-value customization capabilities are often encapsulated in lightweight LoRA modules that are independently trained, distributed, reused, and composed with different base models or additional adapters. Consequently, protecting the base diffusion model or the full generation pipeline does not directly prevent unauthorized copying, redistribution, or reuse of LoRA assets.

Recent studies have started to investigate watermarking mechanisms for parameter-efficient adaptation modules and LoRA assets~\cite{shi2025authenlora,lv2025loraguard,oh2025seal,lin2024efficient,lyu2026lora}. Existing LoRA watermarking methods typically follow a task-centric paradigm, where the ownership signal is injected during the optimization of a specific target LoRA. This paradigm can be abstractly formulated as
\begin{equation}
    \mathcal{L} = \mathcal{L}_{\mathrm{task}} + \lambda \mathcal{L}_{\mathrm{wm}},
\end{equation}
where $\mathcal{L}_{\mathrm{task}}$ preserves the target adaptation behavior and $\mathcal{L}_{\mathrm{wm}}$ enforces watermark recoverability. Under this formulation, watermark embedding and task adaptation are jointly optimized for each individual LoRA. While effective for protecting a single target LoRA, this design tightly couples the ownership signal with a specific adaptation process.
Task-centric LoRA watermarking requires separate watermark-aware training for each asset, which limits scalability and entangles the ownership signal with style or semantic features, reducing robustness under LoRA composition. To address these issues, we adopt a user-centric paradigm where a single reusable Watermark LoRA can be applied to multiple independently trained assets, remaining detectable while preserving generation quality and style fidelity.

\section{Threat Model}
\subsection{User-Centric Watermarking Scenario}

We consider a user-centric LoRA copyright protection scenario, where ownership is associated with the creator rather than with a single LoRA asset. Specifically, an IP owner first holds a user-specific Watermark LoRA that carries a registered secret message. This Watermark LoRA serves as a reusable ownership key and can be attached to different target LoRAs released by the same creator.

Given a target style LoRA, the IP owner protects it by composing the target LoRA with the Watermark LoRA before distribution. In this way, multiple independently trained LoRA assets can share the same ownership signal without requiring watermark retraining for each individual LoRA. The protected LoRA assets may then be released to public communities, commercial platforms, or downstream users, where they can be reused for image generation.

After distribution, a protected LoRA asset may be copied, redistributed, reused with different prompts, merged with other LoRAs, or deployed on third-party generation platforms. The goal of copyright verification is to determine whether a suspicious output is generated from a protected LoRA asset owned by the IP owner.

During verification, the IP owner is given a suspicious generated image or a LoRA-derived output. The owner uses the registered secret message and a private watermark decoder to extract and verify the ownership signal. If the recovered message matches the registered message with sufficiently high confidence, the suspicious output is regarded as originating from the protected LoRA asset.


\subsection{Attacker Capabilities and Goals}

We consider an adversary who has access to the released protected LoRA parameters but does not know the registered secret message or the private watermark decoder. The adversary may attempt to weaken, remove, or obfuscate the embedded ownership signal while maintaining the practical utility of the protected LoRA asset.

Specifically, we consider two categories of attacks. First, the adversary may perform parameter-space manipulations on the released LoRA, including weight pruning, downstream fine-tuning, and merging the protected LoRA with additional LoRA modules. Second, the adversary may apply image-space post-processing to the generated outputs, such as JPEG compression, resizing, cropping, Gaussian noise, blurring, and color perturbation.

The goal of the adversary is to bypass copyright verification and claim unauthorized use of the protected LoRA asset. Such attacks may facilitate unauthorized redistribution, commercial exploitation, or uncredited reuse of the protected LoRA in downstream generation scenarios. However, we assume that the adversary still intends to preserve the usefulness of the LoRA. Therefore, an attack is considered meaningful only if it can suppress watermark detection while largely preserving the visual quality, semantic consistency, and style fidelity of the generated outputs.

Under this threat model, an effective LoRA copyright protection method should satisfy two requirements. First, the embedded ownership signal should remain reliably detectable under common image-space and parameter-space manipulations. Second, the watermark should not noticeably degrade the generation quality or the original adaptation ability of the target LoRA.

\section{LoRA-Key}
\begin{figure*}
    \centering
    \includegraphics[width=1\linewidth]{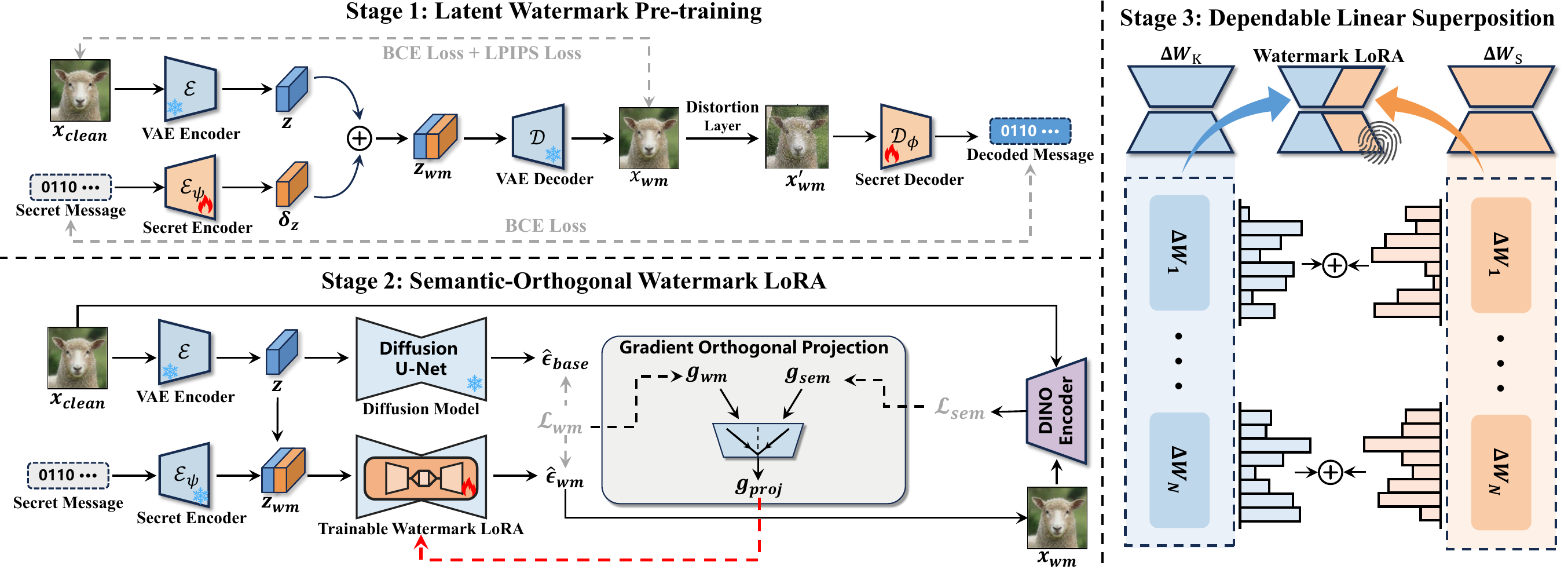}
    \vspace{-5mm}
    \caption{Overview of the proposed LoRA-Key framework. Stage 1 establishes a latent watermark prior for robust message embedding and extraction. Stage 2 learns a semantic-orthogonal Watermark LoRA via watermark consistency learning and Gradient Orthogonal Projection (GOP). Stage 3 enables training-free deployment by linearly superposing the learned Watermark LoRA with a user style LoRA. Ownership is finally verified by recovering the secret message from a suspicious image and performing statistical hypothesis testing.}
    \label{fig:pipeline}
\end{figure*}

In this section, we present the overall pipeline of LoRA-Key, as illustrated in Fig.~\ref{fig:pipeline}. Our method is built upon a modular four-part pipeline: latent watermark prior learning, semantic-orthogonal watermark adaptation, training-free deployment by linear superposition, and ownership verification. Specifically, we first pre-train a robust latent watermark prior for message embedding and extraction. Based on this prior, we then optimize a standalone Watermark LoRA, where watermark fitting is guided by consistency learning and its semantic-conflicting gradient component is explicitly removed via Gradient Orthogonal Projection. After training, the resulting Watermark LoRA can be directly combined with an arbitrary user style LoRA through simple linear superposition, without additional retraining or architectural modification. Finally, for a suspicious image, ownership is determined by recovering the candidate message and conducting hypothesis testing against the registered secret. Through this design, LoRA-Key turns copyright protection into a reusable, plug-and-play module for LoRA-centric diffusion model deployment.

%

\subsection{Latent Watermark Pre-training}
To establish a robust watermark embedding and extraction mechanism in latent space, we first pre-train a private watermark codec using the frozen Variational Autoencoder (VAE) of the base diffusion model, following the general latent watermarking paradigm in prior work~\cite{feng2024aqualora}. Specifically, we jointly optimize a message encoder $\mathcal{E}_{\psi}$ and a private watermark decoder $\mathcal{D}_{\phi}$.

Given a clean reference image $x_{\mathrm{clean}}$, we first obtain its latent representation through the frozen VAE encoder, i.e., $z = \mathcal{E}_{\mathrm{vae}}(x_{\mathrm{clean}})$. For an $L$-bit binary secret message $m \in \{0,1\}^{L}$, the message encoder $\mathcal{E}_{\psi}$ produces a latent residual with the same dimensionality as $z$. This residual is then added to the clean latent to obtain the watermarked latent representation:

\begin{equation}
\begin{aligned}
z_{\mathrm{wm}} = z + \mathcal{E}_{\psi}(m).
\end{aligned}
\end{equation}

The watermarked latent is subsequently decoded by the frozen VAE decoder $\mathcal{D}_{\mathrm{vae}}$ to produce a watermarked image $x_{\mathrm{wm}} = \mathcal{D}_{\mathrm{vae}}(z_{\mathrm{wm}})$.
To improve robustness against the image-level distortions considered in our threat model, we introduce a differentiable distortion layer $\mathcal{N}(\cdot)$ during codec training. This layer simulates common post-processing perturbations, including JPEG compression, Gaussian noise, and random cropping. The distorted image $x'_{\mathrm{wm}} = \mathcal{N}(x_{\mathrm{wm}})$ is re-encoded into latent space by the frozen VAE encoder, and the private decoder predicts the embedded message as

\begin{equation}
\begin{aligned}
m' = \mathcal{D}_{\phi}\big(\mathcal{E}_{\mathrm{vae}}(x'_{\mathrm{wm}})\big).
\end{aligned}
\end{equation}

The encoder $\mathcal{E}_{\psi}$ and decoder $\mathcal{D}_{\phi}$ are jointly optimized through the following objective:
\begin{equation}
\begin{aligned}
\mathcal{L}_{prior} &= \mathcal{L}_{BCE}(m, m') 
+ \lambda_1 \mathcal{L}_{MSE}(x_{clean}, x_{wm}) \\
&\quad + \lambda_2 \mathcal{L}_{LPIPS}(x_{clean}, x_{wm})
\end{aligned}
\end{equation}
where $\mathcal{L}_{\mathrm{BCE}}$ encourages accurate message recovery, while $\mathcal{L}_{\mathrm{MSE}}$ and $\mathcal{L}_{\mathrm{LPIPS}}$ constrain the visual distortion introduced by watermark embedding.
After convergence, we retain and freeze the private decoder $\mathcal{D}_{\phi}$ for the subsequent optimization stage.

\subsection{Semantic-Orthogonal Watermark Injection}
In this stage, we optimize the Watermark LoRA $\Delta \Theta_k$ through a gradient-orthogonalized consistency learning scheme. Instead of treating watermark fitting and semantic preservation as a naively scalarized multi-objective problem, we introduce a Gradient Orthogonal Projection (GOP) mechanism to suppress the component of watermark adaptation that conflicts with image semantics. From a geometric perspective, the core difficulty lies in the non-orthogonal superposition of watermark-oriented updates and semantic-preserving adaptation directions in parameter-update space. When optimization is driven only by watermark reconstruction, the model may bind the watermark signal to content-dependent semantic cues, such as object contours, textures, or local color patterns. Such update-level entanglement introduces conflicting parameter components into the subspace responsible for faithful denoising and semantic consistency, which in turn leads to semantic distortion and degraded robustness under subsequent LoRA composition, merging, or semantic shifts.

To optimize $\Delta \Theta_k$, we use the message-conditioned latent watermark prior established in Stage 1 as a training stimulus. Given a clean latent representation $z$, Stage 1 provides a pre-trained encoder $\mathcal{E}_{\psi}$ to generate a message-dependent residual, which produces a message-dependent residual that is added to $z$ to obtain the watermarked latent representation $z_{wm}$. Noise is then added to both $z$ and $z_{wm}$ at timestep $t$, yielding $z_t$ and $z_{wm,t}$, respectively. The noisy clean latent is fed into the frozen base U-Net to obtain a reference noise prediction $\hat{\epsilon}_{base} = \epsilon_{\theta}(z_t, t, c)$, while the noisy watermarked latent is fed into the U-Net equipped with the trainable Watermark LoRA, yielding $\hat{\epsilon}_{key} = \epsilon_{\theta + \Delta \Theta_k}(z_{wm,t}, t, c)$. We define the watermark consistency objective as the mean squared error between these two predictions:
\begin{equation}
    \mathcal{L}_{wm} = \left\| \hat{\epsilon}_{key} - \hat{\epsilon}_{base} \right\|_2^2.
\end{equation}

This objective encourages the trainable Watermark LoRA to absorb the message-induced latent perturbation into its parameter update, while remaining aligned with the native denoising trajectory of the base model. In other words, the model is encouraged to treat the watermark residual as a structured perturbation that should be compensated by adaptation, rather than by distorting the underlying semantic content.

To constrain semantic distortion at the image level, we further introduce a DINO-based semantic consistency objective. From the current noise prediction $\hat{\epsilon}_{key} $, we estimate the clean latent according to the deterministic diffusion scheduler:
\begin{equation}
    \hat{z}_0 =
\frac{
z_{wm,t} - \sqrt{1-\bar{\alpha}_t}\,\hat{\epsilon}_{key}
}{
\sqrt{\bar{\alpha}_t}
}.
\end{equation}

The estimated latent $\hat{z}_0$ is then decoded by the frozen VAE to obtain the reconstructed image $I_{wm}=\mathcal{D}_{vae}(\hat{z}_{0})$. Meanwhile, the original training image is resized to form the reference image $ I_{ref} $. Let $\mathcal{F}_{dino}(I_{ref})$ and $\mathcal{F}_{dino}(I_{wm})$ denote the corresponding global semantic features. The semantic consistency loss is defined as:
\begin{equation}
    \mathcal{L}_{sem} =
1 -
\frac{
\mathcal{F}_{dino}(I_{ref}) \cdot \mathcal{F}_{dino}(I_{wm})
}{
\left\| \mathcal{F}_{dino}(I_{ref}) \right\|_2
\left\| \mathcal{F}_{dino}(I_{wm}) \right\|_2
}.
\end{equation}

In practice, the reference feature is detached from the computational graph and used as a fixed semantic anchor.

Rather than resolving the tension between watermark fitting and semantic preservation only through scalar loss weighting, we address it explicitly in gradient space. Geometrically, let the watermark objective induce an update direction $g_{wm}$ and the semantic-consistency objective induce an update direction $g_{sem}$. When these two directions are non-orthogonal, $g_{wm}$ contains a conflicting projection onto the semantic-preserving direction, so directly updating parameters with $g_{wm}$ will inevitably perturb the parameter subspace responsible for semantic fidelity. To prevent this first-order interference, we project out the conflicting component before applying the update.
For the U-Net LoRA parameters, denoted by $\Theta_k^{U}$, we independently compute the watermark gradient $\mathbf{g}_{wm}=\nabla_{\Theta_k^{U}} \mathcal{L}_{wm}$ and the semantic gradient $\mathbf{g}_{sem} = \nabla_{\Theta_k^{U}} \mathcal{L}_{sem}$. We then compute the projection coefficient:
\begin{equation}
    \alpha = \frac{\langle \mathbf{g}_{wm}, \mathbf{g}_{sem} \rangle}{\left\| \mathbf{g}_{sem} \right\|_2^2 + \epsilon},
\end{equation}
where $\epsilon$ is a small constant for numerical stability. The projected gradient is given by $\mathbf{g}_{proj} = \mathbf{g}_{wm} - \alpha \mathbf{g}_{sem}$. This projected gradient is then used as the effective update direction for the U-Net LoRA parameters. As a result, the first-order component of the watermark update that conflicts with semantic preservation is explicitly removed, and the resulting adaptation is steered toward the orthogonal complement of the semantic-conflicting direction. Under gradient accumulation, $\mathbf{g}_{proj}$ is added to any previously accumulated gradients before the optimizer step. In the current implementation, the orthogonal projection is applied only to the U-Net LoRA parameters; if text-encoder LoRA parameters are simultaneously trained, they retain the standard watermark-consistency gradient without semantic projection. The watermark-conditioned latent perturbation introduced in this stage is used only during optimization as a stimulus for learning a stable watermark-carrying parameter update. After convergence, the watermark pattern is distilled into $\Delta \Theta_k$, and no additional latent perturbation is required during deployment.

\subsection{Dependable Linear Superposition}
After Stage 2, we obtain the Watermark LoRA $\Delta \Theta_k$, whose update direction has been optimized by GOP to reduce non-orthogonal entanglement with semantic- and style-preserving adaptation directions. In this stage, we describe how this module is fused with a user style LoRA during practical distribution and deployment.

In a typical copyright protection scenario, a creator possesses a frozen base diffusion model $\Theta_0$ and a customized style LoRA $\Delta \Theta_s$ trained for a specific artistic style or concept. To protect this asset, the creator does not need to perform any additional retraining or structural modification on the style LoRA. Instead, before release, the creator directly combines the pre-trained Watermark LoRA with the style LoRA through linear superposition. The deployed model is given by
\begin{equation}
    \Theta_{\mathrm{deploy}} = \Theta_0 + \alpha \Delta \Theta_s + \gamma \Delta \Theta_k,
\end{equation}
where $\alpha$ and $ \gamma $ are the scaling coefficients of the style LoRA and the Watermark LoRA, respectively, controlling their relative strengths in deployment.

The effectiveness of this training-free fusion strategy stems from the geometric optimization property established by GOP. Because the semantic-conflicting component of the watermark gradient has already been projected out during Stage 2, the learned Watermark LoRA is encouraged to occupy a parameter direction that is less entangled with the style-related adaptation encoded in $\Delta \Theta_{s}$. Consequently, the compatibility of $\Delta \Theta_{k}$ with downstream style LoRAs is not enforced post hoc through heuristic fusion rules, but is built directly into the optimization of the Watermark LoRA itself. As a result, directly combining the two LoRA modules introduces substantially less destructive interference than conventional watermarking schemes in which watermark features are entangled with content- or style-dependent semantics.

Consequently, during inference with $ \Theta_{\mathrm{deploy}} $, the model can preserve the generation utility and artistic characteristics of $ \Delta \Theta_s$ to a large extent, while maintaining the watermark-carrying capability introduced by $\Delta \Theta_k$. This makes linear superposition a simple, efficient, and practically dependable deployment strategy: it enables plug-and-play copyright protection without any additional retraining.

\subsection{Ownership Verification}
In practical copyright tracing scenarios, ownership verification can be formulated as a statistical hypothesis testing problem. Let \( m \in \{0,1\}^{L} \) denote the original \( L \)-bit secret message registered by the copyright owner. Given a suspicious image \( I_{\mathrm{sus}} \), the owner first applies the frozen VAE encoder \( \mathcal{E}_{\mathrm{vae}} \) and the watermark decoder \( \mathcal{D}_{\phi} \) to recover a candidate message
\begin{equation}
    m' = \mathcal{D}_{\phi}\big(\mathcal{E}_{\mathrm{vae}}(I_{\mathrm{sus}})\big).
\end{equation}
To determine whether the suspicious image originates from the protected asset, we define the following hypotheses:
\begin{itemize}
    \item \( \mathcal{H}_0 \): The suspicious image \( I_{\mathrm{sus}} \) does not contain the target watermark message \( m \).
    \item \( \mathcal{H}_1 \): The suspicious image \( I_{\mathrm{sus}} \) contains the target watermark message \( m \).
\end{itemize}

Let \( M(m, m') \) denote the number of matching bits between the recovered message \( m' \) and the original message \( m \). We use \( M(m, m') \) as the test statistic for watermark presence. If
\begin{equation}
    M(m, m') > \tau_{\mathrm{ver}},
\end{equation}
where \( \tau_{\mathrm{ver}} \) is a predefined verification threshold, we reject \( \mathcal{H}_0 \) and accept \( \mathcal{H}_1 \), thereby identifying the suspicious image as originating from the protected asset.

To determine \( \tau_{\mathrm{ver}} \), we analyze the distribution of extracted bits under \( \mathcal{H}_0 \). Following a common verification assumption, we model the recovered bits from clean, unwatermarked images as independent Bernoulli random variables with success probability \( 0.5 \). Under this assumption, the matching bit count \( M \) follows a binomial distribution \( \mathcal{B}(L, 0.5) \). The corresponding false positive rate (FPR), i.e., the probability of incorrectly claiming ownership for an unwatermarked image, is given by
\begin{equation}
    \mathrm{FPR}(\tau_{\mathrm{ver}})
    =
    P(M > \tau_{\mathrm{ver}} \mid \mathcal{H}_0)
    =
    \sum_{i=\tau_{\mathrm{ver}}+1}^{L}
    \binom{L}{i}
    \left(\frac{1}{2}\right)^{L}.
\end{equation}
Equivalently, this binomial tail probability can be written using the regularized incomplete beta function as
\begin{equation}
    \mathrm{FPR}(\tau_{\mathrm{ver}})
    =
    I_{1/2}\!\left(\tau_{\mathrm{ver}}+1,\, L-\tau_{\mathrm{ver}}\right).
\end{equation}

In practice, the verification threshold \( \tau_{\mathrm{ver}} \) is selected to keep the false positive rate below a target level. For model-side verification, the owner may generate multiple probe images and average the extraction results across samples, and then report the corresponding true positive rate (TPR) under the same threshold to evaluate the robustness of LoRA-Key.

\section{Experimental Evaluation}
In this section, we first introduce the experimental setup in Section~\ref{sec:setup}. Then, we present the performance evaluation of the proposed LoRA-Key in Section~\ref{sec:eval}, focusing on semantic preservation and watermark robustness. In Section~\ref{sec:generalizability}, we further verify its generalizability across various diffusion models and the effectiveness of the plug-and-play modules. To further analyze the contribution of individual components, we conduct ablation studies in Section~\ref{sec:ablation}.

\subsection{Experimental Setup} \label{sec:setup}

\noindent\textbf{Datasets and Prompts.}
To evaluate LoRA-Key under realistic personalized generation scenarios, we construct both controlled and real-world LoRA evaluation settings. 
In the controlled setting, we train style-driven LoRAs using the StyleDrop dataset~\cite{sohn2023styledrop} and content-driven LoRAs using the DreamBooth dataset~\cite{ruiz2023dreambooth}, covering artistic-style and subject-specific personalization. 
For each target LoRA, we collect or design LoRA-specific seed prompts that reflect its intended usage, such as subject identity, visual attributes, artistic style, and scene composition. 
Based on these seed prompts, we apply GPT-assisted prompt expansion to construct a diverse LoRA-specific prompt pool, from which 3,000 prompts are sampled for each LoRA during evaluation.
The final test images are generated by the corresponding LoRA-adapted diffusion models using these prompts.
We further evaluate LoRA-Key on publicly available community LoRA modules collected from platforms such as Civitai and Hugging Face. 
For each community LoRA, seed prompts are constructed according to its recommended trigger words, usage examples, and target generation domain, and are then expanded into a LoRA-specific prompt pool. 
From each pool, 3,000 prompts are sampled to generate test images with the corresponding community LoRA.

\vspace{0.3em}
\noindent\textbf{Baselines.} To validate the effectiveness of LoRA-Key, we compare it against seven mainstream watermarking methods:
(i) Frequency-based Methods: DwtDct and DwtDctSvd~\cite{cox2007digital}, which employ frequency decomposition as the embedding strategy.
(ii) GAN-based Methods: RivaGAN~\cite{zhang2019robust} and StegaStamp~\cite{tancik2020stegastamp}, which embed watermarks through generative adversarial learning.
(iii) Diffusion-based Methods: Stable Signature~\cite{fernandez2023stable} and Aqualora~\cite{feng2024aqualora}, which embed watermarks directly within the latent space of diffusion models. (iv) LoRA watermarking method: AuthenLoRA~\cite{shi2025authenlora}, which embeds ownership signals during the training of each target LoRA and therefore serves as a task-centric LoRA watermarking baseline. For all baselines, we use the official implementations released by the authors.

\vspace{0.3em}
\noindent\textbf{Evaluation Metrics.} 
Following prior work, we employ metrics across two dimensions: semantic maintenance and watermark preservation.
(i) Semantic Maintenance: To verify visual quality and semantic fidelity, we use the FID~\cite{heusel2017gans} to measure the statistical distance between image distributions, where lower scores indicate more natural results. CLIP Score~\cite{hessel2021clipscore} is introduced to quantify the alignment between generated images and text prompts. Additionally, we employ DreamSim~\cite{fu2023dreamsim} to capture human-level perception of image layout and mid-level semantic consistency; lower values represent higher perceptual similarity to the original.
(ii) Robustness and Detection: We evaluate robustness using Bit Accuracy to quantify the precision of watermark extraction. Based on a preset False Positive Rate (FPR) of $10^{-6}$, we determine the decision threshold and report the True Positive Rate (TPR) to evaluate detection reliability between watermarked and clean images.

\vspace{0.3em}
\noindent\textbf{Robustness Evaluation Settings.} We conduct experiments under a range of distortions:
(i) Common Perturbations: These include resizing (0.5$\times$ downsampling), Gaussian blur (G. blur, 3$\times$3 kernel, $\sigma=4.0$), Gaussian noise (G. noise, 10\% intensity), JPEG compression (quality=50), and color jittering (including brightness, contrast, and saturation adjusted between 0.8 and 1.2), and image sharpening (factor=10.0).
(ii) Adaptive Detection Threshold: Considering varying watermark lengths ($k$-bits), we calculate the minimum successful bits $\tau$ using a Bernoulli distribution under the FPR constraint. An image is identified as watermarked only if the bit accuracy exceeds this threshold ratio, ensuring high reliability and minimal false alarms.

\vspace{0.3em}
\noindent\textbf{Implementation Details.}
Our framework is built upon Stable Diffusion v1.4. The training pipeline consists of two stages: latent watermark prior pre-training and Watermark LoRA optimization. 
In the first stage, we train the encoder-decoder on a 10K-image subset of COCO2014\cite{lin2014microsoft}, where a 48-bit hidden message is encoded into a latent residual $W \in \mathbb{R}^{4 \times 128 \times 128}$ and injected into the diffusion latent space. 
In the second stage, we optimize the watermark LoRA using 3,000 prompts from DiffusionDB\cite{wang2022diffusiondb}. For LoRA adaptation, we fine-tune the projection matrices $W_q$, $W_k$, $W_v$, and $W_{out}$ in all U-Net Transformer blocks with a rank of 64. We optimize the model using AdamW, with learning rates of $1\times10^{-4}$ for the U-Net and $1\times10^{-5}$ for the text encoder. A constant learning rate schedule with 500 warmup steps is adopted. To preserve semantic fidelity, we use a DINOv2 model ~\cite{oquab2023dinov2} as the semantic constraint and apply orthogonal gradient projection to remove the component of the watermark gradient aligned with the semantic gradient direction. 
Fine-tuning is conducted for 15,000 steps with a batch size of 1 and gradient accumulation over 4 steps. During inference, we use the DPM-Solver++ sampler with 50 steps and a guidance scale of 7.5 to generate images at a resolution of $512 \times 512$.

\begin{table*}[t]
\centering
\caption{Comparison of Model Fidelity and Watermark Effectiveness. Bold indicates the best performance in each category.}
\label{tab:results_bold}
\resizebox{0.9\textwidth}{!}{
\begin{tabular}{@{}clccccccc@{}}
\toprule
\multirow{2}{*}{Task} & \multirow{2}{*}{Method} & \multicolumn{3}{c}{Model Fidelity} & \multicolumn{4}{c}{Watermark Effectiveness} \\ \cmidrule(lr){3-5} \cmidrule(lr){6-9}
 &  & FID $\downarrow$ & CLIP $\uparrow$ & DreamSim $\downarrow$ & Bit Acc $\uparrow$ & Bit Acc (Adv.) $\uparrow$ & TPR $\uparrow$ & TPR (Adv.) $\uparrow$ \\ \midrule
\multirow{9}{*}{Content} & None & \textbf{17.53} & \textbf{30.15} & - & - & - & - & - \\
 & DwtDct & 18.26 & 29.91 & \textbf{0.18} & 90.48 & 47.13 & 0.67 & 0.03 \\
 & DwtDctSvd & 19.64 & 29.18 & 0.19 & \textbf{100.00} & 56.88 & \textbf{1.00} & 0.10 \\
 & RivaGAN & 21.76 & 28.82 & 0.19 & \textbf{100.00} & 88.88 & \textbf{1.00} & 0.61 \\
 & StegaStamp & 24.11 & 28.45 & 0.21 & 99.93 & 92.63 & \textbf{1.00} & 0.90 \\
 & Stable Signature & 21.69 & 28.57 & 0.21 & 98.30 & 86.23 & \textbf{1.00} & 0.67 \\ \cmidrule(l){2-9} 
 & AquaLoRA & 20.22 & 24.28 & 0.26 & 86.38 & 85.38 & 0.73 & 0.68 \\
 & AuthenLoRA & 19.85 & 25.10 & 0.24 & 94.27 & 92.15 & 0.91 & 0.84 \\
 & Ours & 18.72 & 29.85 & 0.20 & 98.23 & \textbf{97.85} & \textbf{1.00} & \textbf{1.00} \\ \midrule
\multirow{9}{*}{Style} & None & \textbf{18.12} & \textbf{27.15} & - & - & - & - & - \\
 & DwtDct & 18.65 & 26.95 & \textbf{0.17} & 90.14 & 47.25 & 0.64 & 0.01 \\
 & DwtDctSvd & 19.66 & 26.78 & \textbf{0.17} & \textbf{99.99} & 56.50 & \textbf{1.00} & 0.10 \\
 & RivaGAN & 21.22 & 26.55 & 0.18 & 99.89 & 88.00 & 0.99 & 0.59 \\
 & StegaStamp & 33.61 & 26.12 & 0.20 & 99.86 & 93.50 & \textbf{1.00} & 0.94 \\
 & Stable Signature & 27.20 & 25.88 & 0.21 & 99.43 & 87.99 & \textbf{1.00} & 0.72 \\ \cmidrule(l){2-9} 
 & AquaLoRA & 31.42 & 21.45 & 0.28 & 86.76 & 85.22 & 0.75 & 0.67 \\
 & AuthenLoRA & 28.90 & 22.80 & 0.25 & 94.93 & 91.04 & 0.93 & 0.81 \\
 & Ours & 20.55 & 26.80 & 0.19 & 99.96 & \textbf{99.71} & \textbf{1.00} & \textbf{1.00} \\ \bottomrule
\end{tabular}
}
\end{table*}

\subsection{Semantic Preservation and Watermark Robustness} \label{sec:eval}
\noindent\textbf{Semantic Preservation.}
We evaluate semantic preservation using three complementary metrics: FID for distribution similarity, CLIP for semantic alignment, and DreamSim for perceptual consistency. 
As shown in Table~\ref{tab:results_bold}, LoRA-Key maintains competitive model fidelity among all watermarking baselines, while showing clear advantages over LoRA-based watermarking methods.
In content-driven tasks, LoRA-Key achieves an FID of 18.72 and a CLIP score of 29.85, remaining close to the clean model. 
Compared with conventional image-level watermarking methods, LoRA-Key avoids the severe fidelity degradation observed in methods such as RivaGAN and StegaStamp, whose FID scores increase to 21.76 and 24.11, respectively. 
Compared with diffusion- and LoRA-based baselines, LoRA-Key also shows stronger semantic preservation: it reduces FID by 2.97 over Stable Signature, 1.50 over AquaLoRA, and 1.13 over AuthenLoRA, while maintaining a high CLIP score comparable to the clean model. 
These results indicate that LoRA-Key better preserves target content semantics and prompt-relevant visual attributes after watermark attachment.
This improvement mainly comes from the reduced interference between watermark embedding and target LoRA adaptation. 
AquaLoRA is designed to protect the diffusion U-Net itself, and its watermark behavior may interfere with an additional target LoRA during composition. 
AuthenLoRA embeds ownership signals during the training of each target LoRA, which couples watermark learning with task-specific content or style adaptation. 
In contrast, LoRA-Key encapsulates the ownership signal into an independent Watermark LoRA and uses GOP to remove the semantic-conflicting component of the watermark update, thereby reducing parameter interference during modular composition.
In style-driven tasks, LoRA-Key also achieves strong semantic and perceptual preservation. 
Compared with neural and diffusion-based watermarking methods, LoRA-Key avoids the severe quality degradation observed in StegaStamp, Stable Signature, and AquaLoRA, whose FID scores increase to 33.61, 27.20, and 31.42, respectively. 
Among LoRA-related baselines, LoRA-Key reduces FID by 10.87 over AquaLoRA and 8.35 over AuthenLoRA, and improves CLIP by 5.35 and 4.00, respectively. 
It also achieves a lower DreamSim score than AquaLoRA and AuthenLoRA, indicating better perceptual consistency with the target style. 
These results suggest that LoRA-Key can better preserve style fidelity and global visual structure when composed with style-driven LoRAs.

\begin{figure}[t]
\centering
  \includegraphics[width=\columnwidth]{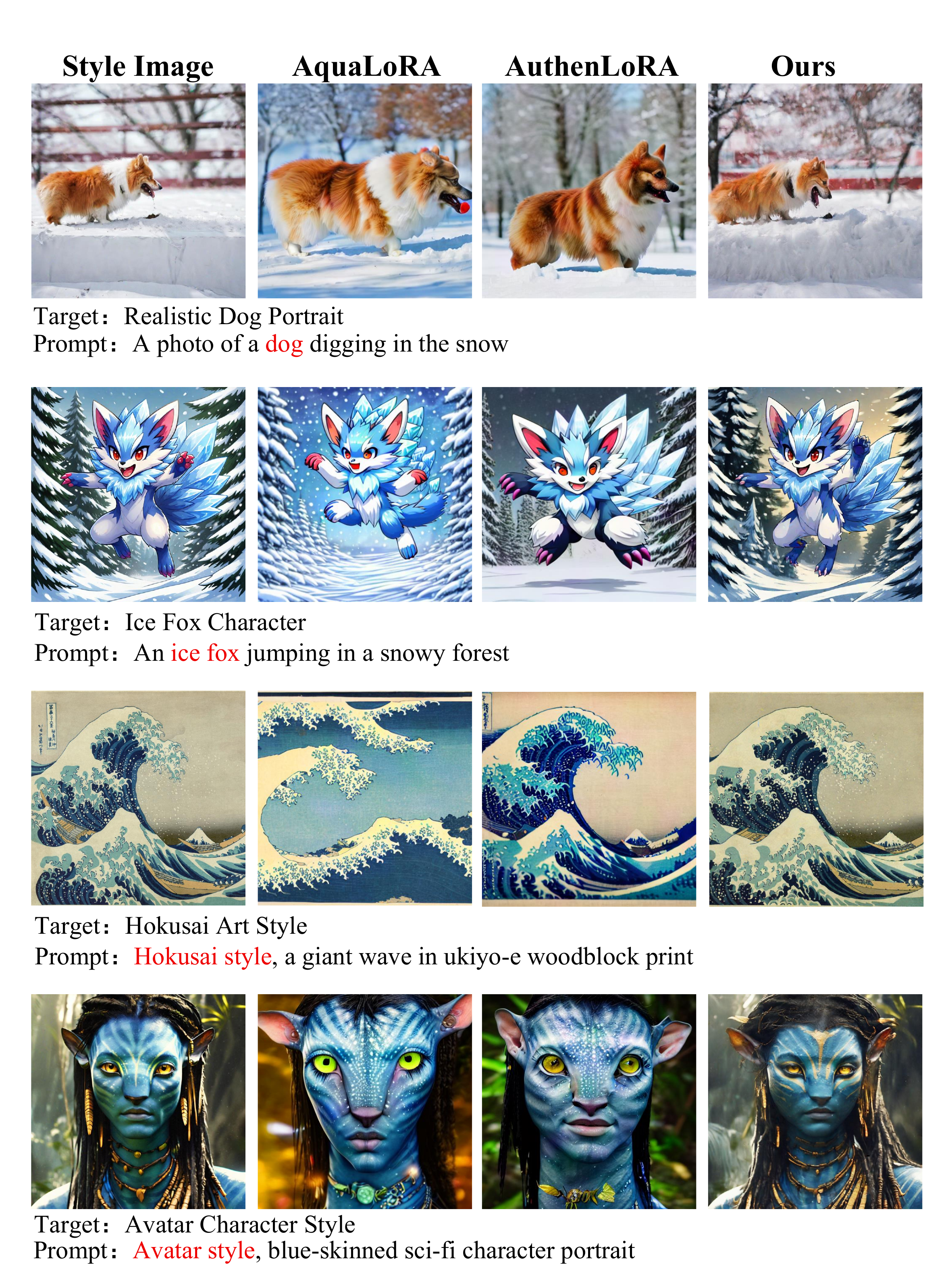}
    \caption{Qualitative comparison of semantic preservation under content-driven and style-driven generation. The reference image in each row denotes the corresponding exemplar for LoRA fine-tuning.}
    \label{fig:semantic_vis}
\end{figure}

Beyond quantitative metrics, we further provide qualitative comparisons in Fig.\ref{fig:semantic_vis} to visually assess semantic preservation under both content-driven and style-driven settings. 
The results show that LoRA-Key better preserves the core content semantics and prompt-relevant visual attributes when the watermark module is composed with a target LoRA.
In the content-driven examples, such as ``a photo of a dog digging in the snow'' and ``an ice fox jumping in a snowy forest,'' AquaLoRA and AuthenLoRA exhibit semantic deviation to different degrees, including inconsistent pose, weakened action semantics, and distorted local structures. 
This is mainly because their watermark signals are more tightly coupled with the protected model or the target LoRA training process, which can introduce interference during LoRA composition. 
By contrast, LoRA-Key better preserves the content identity, action semantics, and scene coherence, producing images that remain visually closer to the target LoRA outputs while still carrying the watermark.
In the style-driven examples, such as Hokusai-style and Avatar-style generation, AquaLoRA and AuthenLoRA tend to introduce style-semantic mismatch or structural instability. 
For example, the generated results may deviate from the target wave structure or weaken the intended character style. 
In comparison, LoRA-Key maintains more consistent global composition and clearer semantic structure while better respecting the target artistic style.

\vspace{0.3em}
\noindent\textbf{Watermark Robustness.} 
In practical deployment, generated images are often subjected to a variety of post-processing distortions that may weaken or erase embedded copyright signals. 
As summarized in Table~\ref{tab:robustness_simplified}, we conduct a comprehensive robustness evaluation under eight common perturbations, comparing LoRA-Key with state-of-the-art baselines in both content-driven and style-driven settings. 
The results reveal clear limitations of existing watermarking paradigms. 
Traditional frequency-domain methods such as DwtDct and DwtDctSvd remain close to random guessing under many distortions, with bit accuracy around 50\%, indicating that their shallow embedding mechanisms are highly vulnerable to post-processing perturbations. 
GAN-based approaches such as StegaStamp show stronger robustness, but their performance still deteriorates under certain distortions; for instance, the bit accuracy drops to 77.10\% under Gaussian blur. 
Diffusion-based methods, including Stable Signature and AquaLoRA, also show limited resilience to stochastic corruption, with Stable Signature falling to 56.07\% under Gaussian noise. 
AuthenLoRA embedding ownership signals during target LoRA training, achieving average bit accuracy of 92.15\% on Content tasks and 91.04\% on Style tasks. 
However, its robustness remains lower than LoRA-Key under all evaluated perturbations, suggesting that jointly optimizing watermark embedding with each target LoRA still entangles the ownership signal with task-specific adaptation and makes it less stable under image-level distortions.
By contrast, LoRA-Key consistently achieves the strongest robustness among all compared methods, reaching an average bit accuracy of 97.85\% on Content tasks and 99.71\% on Style tasks. 
Even under Gaussian noise with an intensity of 10\%, LoRA-Key still maintains bit accuracy of 95.77\% on Content tasks and 98.25\% on Style tasks, clearly outperforming AuthenLoRA, which drops to 86.70\% and 84.15\%, respectively. 
This advantage is largely attributable to the proposed reusable Watermark LoRA and GOP strategy. 
Rather than binding watermark information to fragile image-space artifacts or directly coupling it with each target LoRA training process, LoRA-Key distills the ownership signal into a dedicated low-rank parameter update. 
GOP further removes the semantic-conflicting component of the watermark update, making the learned watermark subspace more stable under image-level distortions and more compatible with different content- and style-driven LoRA adaptations. 
These results demonstrate that LoRA-Key provides robust ownership verification across different personalization regimes.

\begin{table*}[t]
\centering
\caption{Robustness Experiment (Bit Acc). Bold indicates the best performance.}
\label{tab:robustness_simplified}

\setlength{\tabcolsep}{6pt} 
\resizebox{0.9\textwidth}{!}{
\begin{tabular}{@{}clccccccccc@{}}
\toprule
Task & Method & Resize & G. blur & G. noise & JPEG & Bright. & Contrast & Satur. & Sharp. & Avg \\ \midrule
\multirow{8}{*}{Content} 
 & DwtDct & 47.82 & 48.15 & 40.33 & 49.12 & 43.94 & 49.27 & 31.18 & 67.19 & 47.13 \\
 & DwtDctSvd & 49.52 & 56.02 & 72.88 & 50.98 & 43.77 & 46.09 & 42.32 & 93.42 & 56.88 \\
 & RivaGAN & 82.97 & 69.00 & \textbf{100.00} & 88.52 & 73.26 & \textbf{98.27} & \textbf{100.00} & 98.98 & 88.88 \\
 & StegaStamp & \textbf{99.96} & 77.10 & 99.46 & \textbf{100.00} & 72.11 & 93.09 & 99.58 & \textbf{99.70} & 92.63 \\
 & Stable Signature & 60.12 & 87.24 & 56.07 & 81.50 & 97.88 & 98.15 & 98.26 & 98.92 & 84.77 \\ \cmidrule(l){2-11} 
 & AquaLoRA & 86.31 & 86.33 & 81.67 & 84.98 & 85.17 & 86.01 & 86.32 & 86.32 & 85.39 \\
 & AuthenLoRA & 88.45 & 89.12 & 86.70 & 91.35 & 95.80 & 95.10 & 95.95 & 94.73 & 92.15 \\
 & \textbf{Ours} & 98.20 & \textbf{98.21} & 95.77 & 97.89 & \textbf{98.14} & 98.21 & 98.22 & 98.18 & \textbf{97.85} \\ \midrule
\multirow{8}{*}{Style} 
 & DwtDct & 48.10 & 48.75 & 43.52 & 48.22 & 42.42 & 43.67 & 63.17 & 40.15 & 47.25 \\
 & DwtDctSvd & 51.98 & 57.75 & 61.97 & 53.10 & 43.04 & 45.13 & 48.85 & 90.18 & 56.50 \\
 & RivaGAN & 80.41 & 62.87 & 97.08 & 90.15 & 80.83 & 93.57 & 98.98 & \textbf{100.00} & 87.99 \\
 & StegaStamp & \textbf{100.00} & 76.10 & \textbf{99.07} & 99.60 & 88.35 & 89.38 & 95.39 & \textbf{100.00} & 93.49 \\
 & Stable Signature & 60.83 & 91.15 & 68.83 & 85.96 & 99.13 & 99.39 & 99.43 & 99.23 & 87.99 \\ \cmidrule(l){2-11} 
 & AquaLoRA & 86.77 & 86.73 & 81.08 & 84.58 & 83.74 & 86.42 & 86.51 & 85.96 & 85.22 \\
 & AuthenLoRA & 87.20 & 88.54 & 84.15 & 89.92 & 94.88 & 94.50 & 95.12 & 94.01 & 91.04 \\
 & \textbf{Ours} & 99.95 & \textbf{99.95} & 98.25 & \textbf{99.85} & \textbf{99.88} & \textbf{99.94} & \textbf{99.95} & 99.94 & \textbf{99.71} \\ \bottomrule
\end{tabular}
}
\end{table*}

\vspace{0.3em}
\subsection{Scalability and Efficiency}
Scalability is a key advantage of LoRA-Key in user-centric LoRA copyright protection, where a single creator may continuously release multiple independently trained LoRA assets. 
In this scenario, an ideal protection method should avoid repeated watermark training for every newly released LoRA while maintaining reliable ownership verification performance. 
As shown in Table~\ref{tab:scalability_time}, LoRA-Key only needs to train a user-specific Watermark LoRA once, after which the same watermark module can be attached to multiple target LoRAs with negligible additional cost. 
When the number of protected LoRAs increases from 1 to 10, the watermark training time remains fixed at 4.82 hours, while the attachment time increases only from 0.03 to 0.28 hours. 
As a result, the total protection time remains nearly unchanged, increasing from 4.85 to 5.10 hours, and the amortized cost per LoRA decreases from 4.85 to 0.51 hours. 
Meanwhile, LoRA-Key maintains consistently strong verification performance: the TPR remains 1.00, 0.99, and 0.99 for 1, 5, and 10 protected LoRAs, respectively, while the adversarial TPR also remains at 1.00, 0.99, and 0.99.

In contrast, AuthenLoRA follows a task-centric protection paradigm and requires independent watermark-aware training for each target LoRA. 
Its total protection time therefore grows linearly with the number of released LoRA assets, increasing from 15.37 hours for one LoRA to 153.70 hours for ten LoRAs, while the average cost per LoRA remains 15.37 hours. 
Moreover, AuthenLoRA provides lower ownership verification performance under the same scalability setting, with TPR values of 0.93, 0.91, and 0.91 and adversarial TPR values of 0.81, 0.80, and 0.81 for 1, 5, and 10 protected LoRAs, respectively.
This comparison shows that LoRA-Key converts LoRA copyright protection from repeated per-asset retraining into one-time ownership-key training plus lightweight attachment, achieving both higher verification reliability and substantially better protection-time scalability. 
This makes it more suitable for creators who continuously publish and protect growing LoRA repositories.


\begin{table*}[t]
\centering
\caption{Scalability comparison between LoRA-Key and AuthenLoRA with different numbers of protected LoRA modules. Time is reported in hours.}
\label{tab:scalability_time}
\setlength{\tabcolsep}{3.0pt}
\renewcommand{\arraystretch}{0.95}
\footnotesize
\begin{tabular}{c c c c c c c c}
\toprule
Method & LoRAs & TPR & TPR (Adv.) & Watermark Training Time & Watermark Attachment Time & Total Time & Avg./LoRA \\
\midrule
\multirow{3}{*}{LoRA-Key}
& 1  & \textbf{1.00} & \textbf{1.00} & 4.82 & 0.03 & \textbf{4.85} & \textbf{4.85} \\
& 5  & \textbf{0.99} & \textbf{0.99} & 4.82 & 0.15 & \textbf{4.97} & \textbf{0.99} \\
& 10 & \textbf{0.99} & \textbf{0.99} & 4.82 & 0.28 & \textbf{5.10} & \textbf{0.51} \\
\midrule
\multirow{3}{*}{AuthenLoRA}
& 1  & 0.93 & 0.81 & 15.37  & -- & 15.37  & 15.37 \\
& 5  & 0.91 & 0.80 & 76.85  & -- & 76.85  & 15.37 \\
& 10 & 0.91 & 0.81 & 153.70 & -- & 153.70 & 15.37 \\
\bottomrule
\end{tabular}
\end{table*}

\subsection{Generalization and Transferability}\label{sec:generalizability}
\noindent\textbf{Cross-Architecture Generalization.}
To evaluate the backbone-level generalizability of LoRA-Key, we extend our experiments to a diverse set of diffusion architectures, including the classical U-Net-based SD 1.4, the high-resolution dual-encoder model SDXL, and the Transformer-based PixArt-$\alpha$. As summarized in Table~\ref{tab:cross_arch_unified}, LoRA-Key consistently achieves strong performance across all evaluated backbones, indicating that the proposed GOP-based watermark optimization is not restricted to a specific diffusion architecture.
On SDXL, which operates at a substantially higher image resolution of $1024 \times 1024$, LoRA-Key maintains strong visual quality, achieving an FID of 17.72 and a DreamSim score of 0.12. Despite the increased complexity introduced by the larger model scale and dual-text-encoder design, our method still attains a Bit Accuracy of 96.27\%, demonstrating that the learned watermark signal remains stable in a higher-dimensional generative setting.
More importantly, the results on PixArt-$\alpha$ provide further evidence of the architectural flexibility of our framework. Unlike SD\~1.4 and SDXL, PixArt-$\alpha$ is built on a DiT-style Transformer backbone rather than a convolutional U-Net. Nevertheless, LoRA-Key achieves a Bit Accuracy of 99.24\% and a TPR of 1.00 on this model, suggesting that its effectiveness does not depend on backbone-specific spatial priors. This robustness across heterogeneous architectures can be attributed to the fact that GOP operates directly in the parameter-update space of LoRA modules, rather than relying on architecture-specific latent feature patterns. By optimizing watermark-carrying low-rank updates at the weight level, LoRA-Key can be more naturally transferred across backbones with different computational primitives, such as convolutions and self-attention.
These results demonstrate that LoRA-Key generalizes well across representative diffusion architectures, providing a practical foundation for copyright protection in LoRA-centric generative model ecosystems.

\begin{table*}[t]
\centering
\caption{Cross-architecture evaluation of LoRA-Key on representative diffusion backbones in terms of model fidelity and watermark effectiveness.}
\label{tab:cross_arch_unified}
\small
\resizebox{0.9\textwidth}{!}{
\begin{tabular}{@{}l ccc cccc@{}}
\toprule
\multirow{2}{*}{Model} & \multicolumn{3}{c}{Model Fidelity} & \multicolumn{4}{c}{Watermark Effectiveness} \\ \cmidrule(lr){2-4} \cmidrule(lr){5-8}
 & FID $\downarrow$ & CLIP $\uparrow$ & DreamSim $\downarrow$ & Bit Acc $\uparrow$ & Bit Acc (Adv.) $\uparrow$ & TPR $\uparrow$ & TPR (Adv.) $\uparrow$ \\ \midrule
SD 1.4 & 18.72 & 29.85 & 0.20 & 98.23 & 97.86 & 1.00 & 1.00 \\
SDXL & 17.72 & 29.89 & 0.12 & 96.27 & 91.88 & 1.00 & 0.87 \\
PixArt-$\alpha$ & 19.49 & 27.82 & 0.14 & 99.24 & 99.09 & 1.00 & 1.00 \\ \bottomrule
\end{tabular}
}
\end{table*}

\vspace{0.3em}
\noindent\textbf{Effectiveness on Fine-Tuned Models.}
To evaluate the effectiveness of LoRA-Key under downstream fine-tuning, we further test it on several representative models obtained by fine-tuning SD~1.4, including Realistic Vision, ChilloutMix, and DreamShaper. As reported in Table~\ref{tab:transferability_simple}, LoRA-Key maintains strong watermark detectability across all evaluated fine-tuned models, indicating that the learned watermark signal can be reliably preserved even after subsequent model adaptation. In the clean setting, the Bit Accuracy remains above 90\% for all three models, reaching 93.99\% on ChilloutMix, while the corresponding TPR values also stay consistently high. These results suggest that the watermark carried by LoRA-Key is not narrowly tied to the original SD~1.4 parameter configuration, but instead remains effective after downstream fine-tuning.

This trend also persists under adversarial perturbations. As shown in Table~\ref{tab:transferability_simple}, the adversarial Bit Accuracy exhibits only a limited degradation across all evaluated models, while TPR (Adv.) remains above 88\%. For example, LoRA-Key achieves 89.74\% adversarial Bit Accuracy and a TPR (Adv.) of 0.88 on Realistic Vision, and retains 92.81\% adversarial Bit Accuracy with a TPR (Adv.) of 0.93 on ChilloutMix. By preserving reliable ownership verification capability not only on the original base model but also across downstream fine-tuned variants, LoRA-Key confirms its effectiveness and robustness in subsequent tuning tasks.

\vspace{0.3em}
\noindent\textbf{Real-World LoRA.}
To evaluate the practical effectiveness of LoRA-Key in realistic community deployment scenarios, we further test it on publicly available LoRA modules collected from platforms such as Civitai and Hugging Face. As shown in Table~\ref{tab:real_world_dual_tag}, these LoRAs cover six representative domains, including character, portrait, pixel-art, clothing, architecture, and material-oriented generation. Since these modules are trained by different creators with diverse datasets and configurations, they provide a heterogeneous benchmark for assessing cross-LoRA generalizability. The results show that LoRA-Key can be reliably integrated with third-party LoRAs while largely preserving their original generative characteristics, achieving an average FID of 28.38, CLIP score of 27.56, and DreamSim of 0.23. Although a moderate performance drop is observed compared with controlled settings, the method maintains stable watermark detectability across all evaluated domains, with an average bit accuracy of 96.14\% and an adversarial bit accuracy of 94.52\%. These results demonstrate the robustness and practical applicability of LoRA-Key for protecting community-developed LoRA assets.

\begin{table}[t]
\centering
\caption{Cross-domain performance on representative community-developed LoRA models.}
\label{tab:real_world_dual_tag}
\small
\resizebox{\columnwidth}{!}{
\begin{tabular}{@{}l ccc cc@{}}
\toprule
\multirow{2}{*}{\textbf{LoRA Style}} & \multicolumn{3}{c}{\textbf{Image Fidelity}} & \multicolumn{2}{c}{\textbf{Watermark Reliability}} \\ \cmidrule(lr){2-4} \cmidrule(lr){5-6}
 & FID $\downarrow$ & CLIP $\uparrow$ & DreamSim $\downarrow$ & Bit Acc. $\uparrow$ & Bit Acc. (Adv.) $\uparrow$ \\ \midrule
Anime Character & 27.41 & 28.15 & 0.18 & 94.63 & 92.11 \\
Realistic Portrait & 25.93 & 29.42 & 0.20 & 97.12 & 95.48  \\
Pixel Art Style & 29.84 & 28.95 & 0.19 & 97.35 & 95.76 \\
Clothing Design & 26.21 & 26.84 & 0.26 & 93.92 & 92.84 \\
Architecture & 30.77 & 26.12 & 0.28 & 97.88 & 96.63 \\
Crystalline Material & 30.12 & 25.90 & 0.29 & 95.94 & 94.32 \\
 \midrule
\rowcolor[gray]{0.95} Average &28.38 & 27.56 & 0.23 & 96.14 & 94.52\\
\bottomrule
\end{tabular}
}
\end{table}

\begin{table}[t]
\centering
\caption{Effectiveness of LoRA-Key on downstream fine-tuned models derived from SD 1.4 in terms of clean and adversarial watermark detection performance.}
\label{tab:transferability_simple}
\setlength{\tabcolsep}{6pt}
\begin{tabular}{lcccc}
\toprule
Model 
& Bit Acc 
& Bit Acc (Adv.) 
& TPR 
& TPR (Adv.) \\
\midrule
SD 1.4           & 98.23 & 97.85 & 1.00 & 0.99 \\
Realistic Vision & 90.67 & 89.74 & 0.92 & 0.88 \\
ChilloutMix      & 93.99 & 92.81 & 0.97 & 0.93 \\
DreamShaper      & 91.17 & 89.81 & 0.97 & 0.90 \\
\bottomrule
\end{tabular}
\end{table}

\subsection{Ablation Study}  \label{sec:ablation}
\begin{figure}[t]
\centering
\includegraphics[width=\columnwidth]{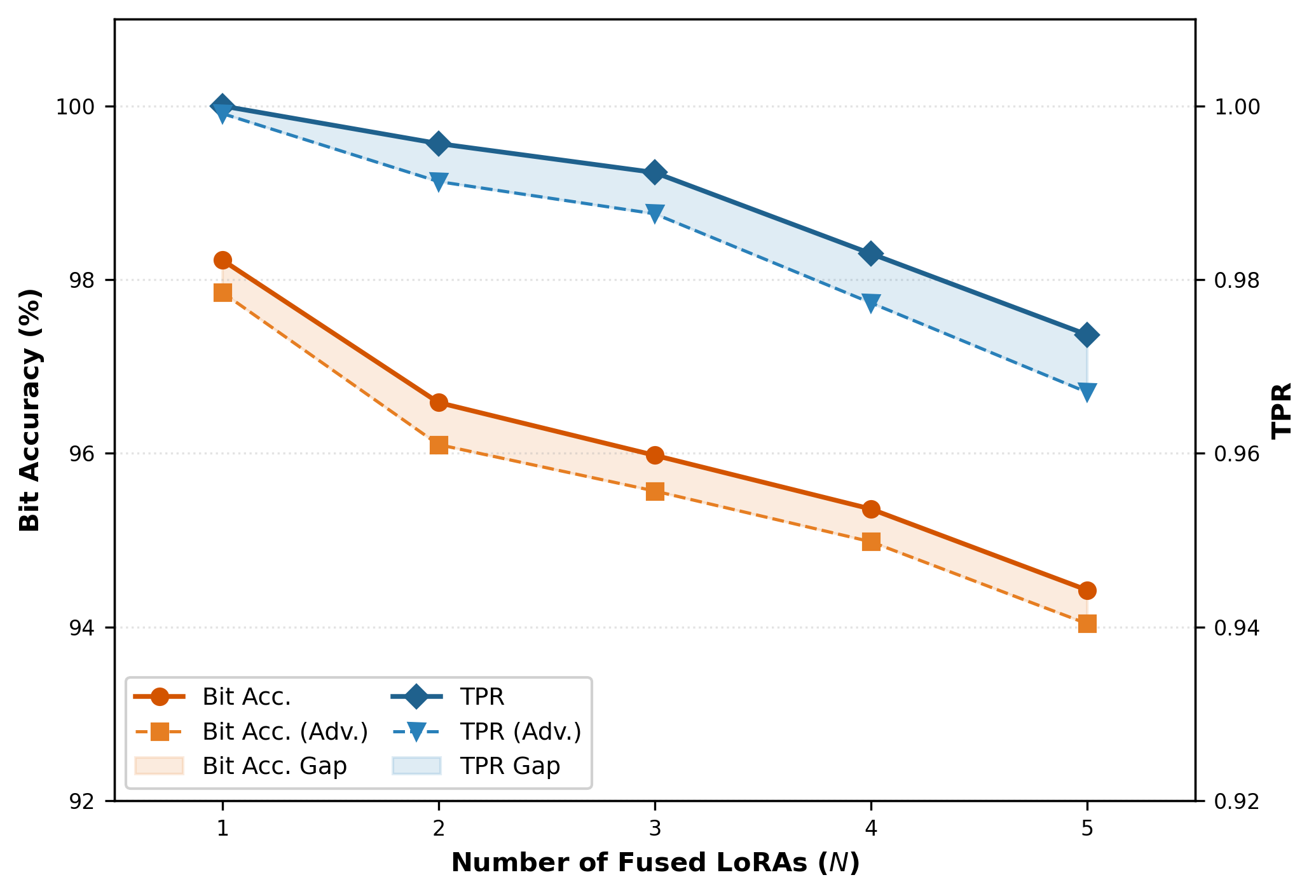}
\caption{Effect of multi-LoRA fusion on watermark extraction performance, reported in terms of Bit Accuracy and TPR as the number of fused LoRA adapters increases from 1 to 5.}
\label{fig:fusion_impact}
\end{figure}

\noindent\textbf{Impact of Multi-LoRA Fusion.}
In practical generation pipelines, multiple LoRA modules are often fused simultaneously to achieve richer stylistic and structural control. To evaluate the robustness of LoRA-Key under such multi-LoRA composition, we progressively increase the number of fused adapters from $N=1$ to $N=5$ during inference. As shown in Fig.~\ref{fig:fusion_impact}, both Bit Accuracy and TPR remain consistently high as the number of fused LoRAs increases, under both clean and adversarial settings.

Although composing more LoRA modules inevitably introduces stronger interactions in the parameter space, these interactions only lead to a gradual performance decline rather than catastrophic watermark failure. Specifically, the clean Bit Accuracy decreases from 98.23\% to 94.44\%, while the adversarial Bit Accuracy drops from 97.85\% to 94.05\% as $N$ increases from 1 to 5. Similarly, TPR remains close to 1.00 throughout, with the adversarial TPR staying above 0.96 even at the largest fusion scale. These results indicate that the watermark signal learned by LoRA-Key is highly resilient to interference from additional LoRA modules, thereby supporting reliable ownership verification in realistic multi-LoRA deployment scenarios.

\noindent\textbf{Robustness to LoRA Rank Variations}
\label{sec:rank_analysis}
To evaluate whether LoRA-Key remains effective when composed with LoRA modules of different capacities, we vary the rank of the style LoRA from 32 to 320 while keeping the watermarked LoRA fixed at rank 64. This setting reflects a practical plug-and-play scenario in which a standardized watermark module needs to remain compatible with downstream LoRAs of heterogeneous parameter budgets.

As reported in Table~\ref{tab:lora_rank}, LoRA-Key maintains strong watermark effectiveness across a wide range of Content LoRA ranks. In particular, Bit Accuracy remains above 94\% in all evaluated cases, and the clean TPR stays at or near 1.00 throughout. The best overall performance is achieved when the host Content LoRA has rank 64, where LoRA-Key reaches an FID of 18.72, a CLIP score of 29.85, and a Bit Accuracy of 98.23\%, while also preserving perfect TPR under both clean and adversarial settings. Even when the style LoRA rank is increased to 320, watermark recovery remains reliable, with Bit Accuracy and adversarial Bit Accuracy still reaching 95.80\% and 94.41\%, respectively.

\begin{table*}[t]
\centering
\caption{Impact of style LoRA rank on model fidelity and watermark effectiveness.  The watermarked LoRA is fixed at rank 64, while the rank of the style LoRA varies from 32 to 320.}
\label{tab:lora_rank}
\resizebox{0.8\textwidth}{!}{
\begin{tabular}{c ccc cccc}
\toprule
\multirow{2}{*}{Rank} & \multicolumn{3}{c}{Model Fidelity} & \multicolumn{4}{c}{Watermark Effectiveness} \\ \cmidrule(lr){2-4} \cmidrule(lr){5-8}
 & FID $\downarrow$ & CLIP $\uparrow$ & DreamSim $\downarrow$ & Bit Acc $\uparrow$ & Bit Acc (Adv.) $\uparrow$ & TPR $\uparrow$ & TPR (Adv.) $\uparrow$ \\ \midrule
320 & 23.47 & \textbf{30.39} & 0.22 & 95.80& 94.41& 0.99 & 0.96 \\
160 & 25.18 & 26.50 & 0.20 & 94.97 & 92.87 & 0.97 & 0.93 \\
80  & 21.93 & 29.31 & 0.20 & 97.32 & 96.89 & 1.00 & 1.00 \\
64  & \textbf{18.72} & 29.85 & \textbf{0.20} & \textbf{98.23} & \textbf{97.85} & \textbf{1.00} & \textbf{1.00} \\
32  & 29.50 & 29.91 & 0.22 & 95.18 & 94.28 & 0.99 & 0.97 \\ \bottomrule
\end{tabular}
}
\end{table*}

\begin{figure}[t]
    \centering
    \includegraphics[width=\columnwidth]{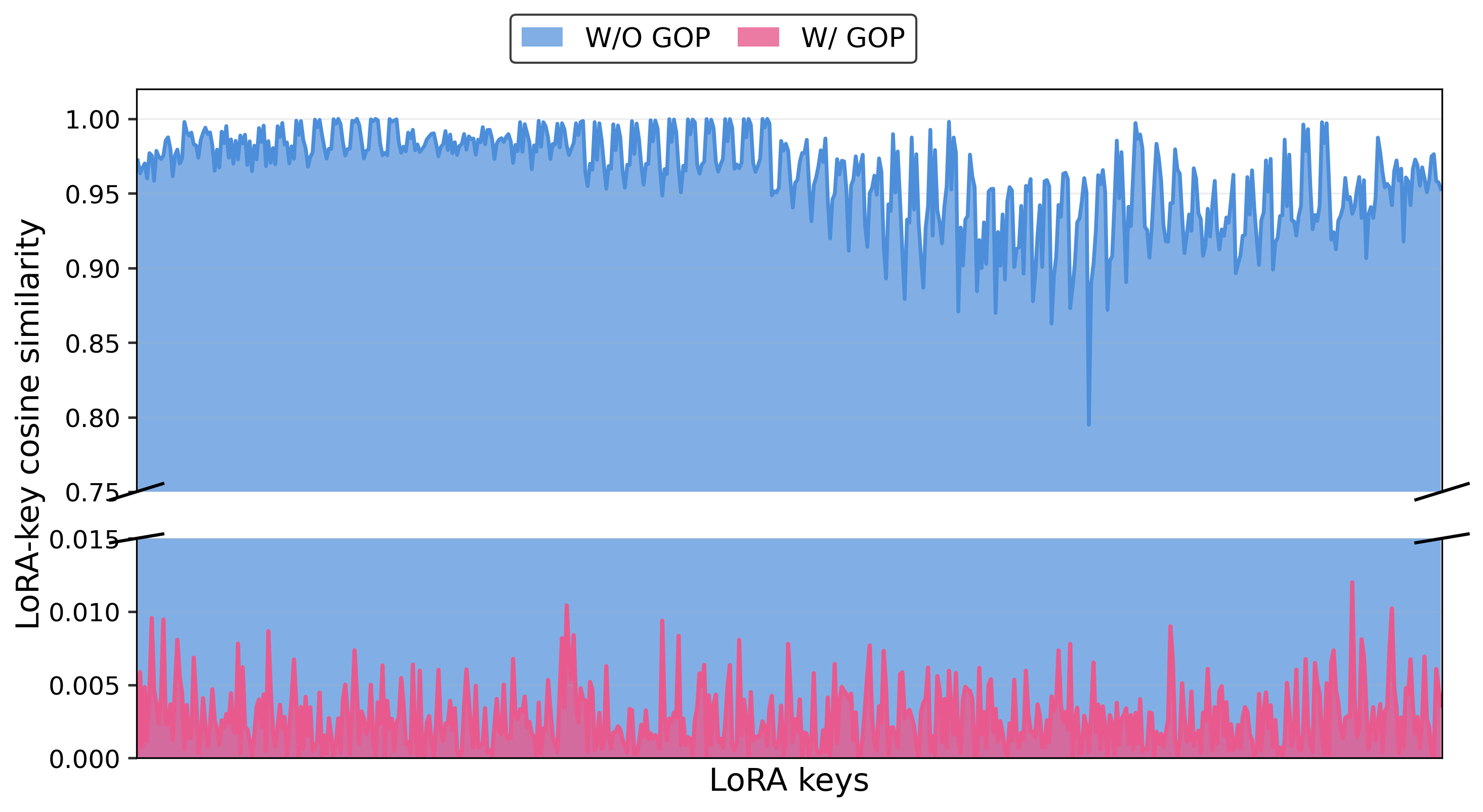} 
 \caption{Effect of GOP on LoRA-Key cosine similarity. Without GOP, cosine similarities remain high and cluster near 1, indicating strong overlap between LoRA updates. With GOP, the similarities are sharply suppressed toward zero. A broken y-axis is used to highlight the contrast between the two regimes.}
    \label{fig:transferability_vis}
\end{figure}

\begin{figure}[t]
    \centering
 \includegraphics[width=\columnwidth]{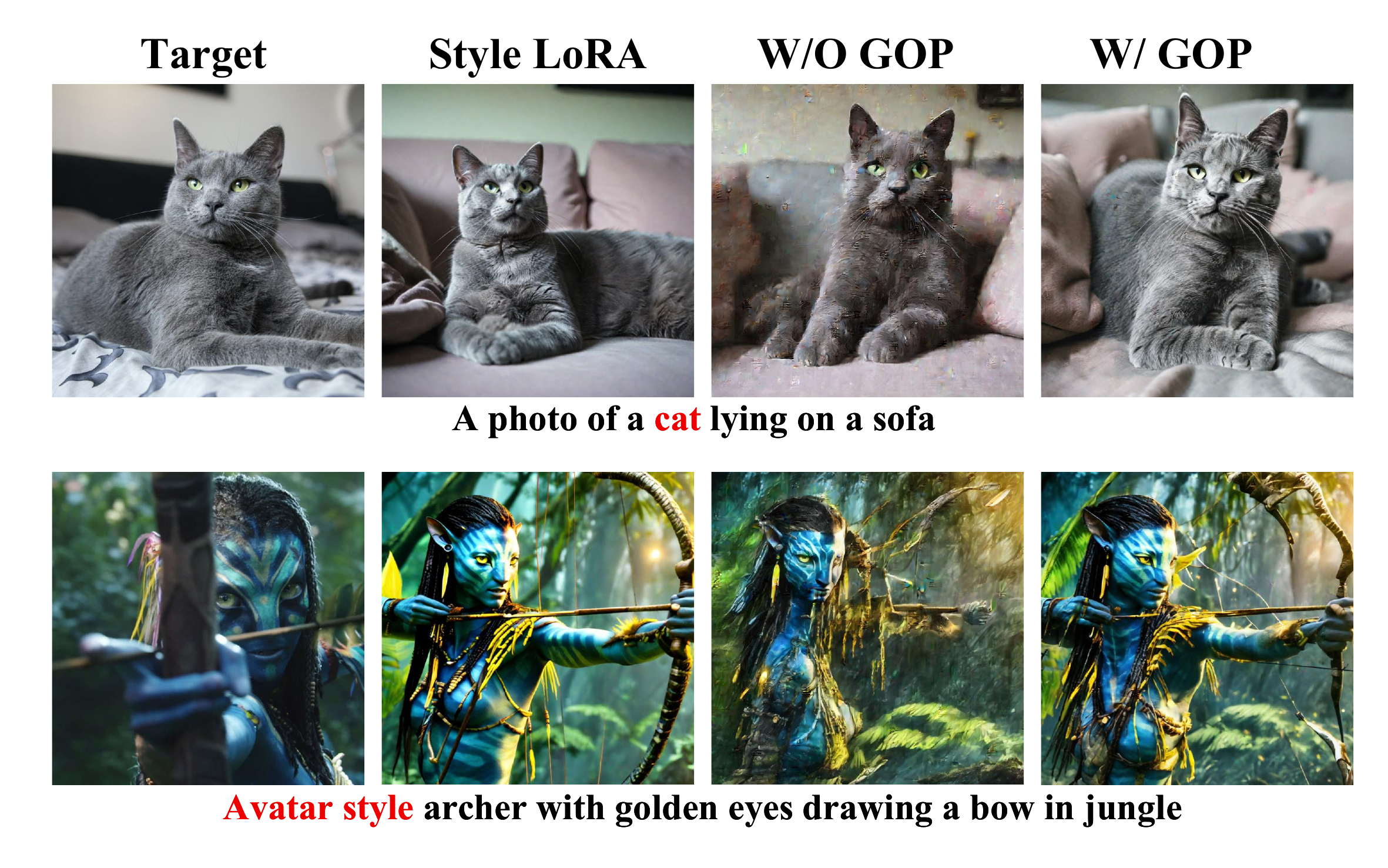} 
    
    \caption{Visual ablation of the GOP mechanism. W/O GOP, composing the watermark module with a style LoRA leads to noticeable degradation in fidelity and style consistency. W/ GOP, the generated results better preserve subject structure and target stylistic characteristics.}
    \label{fig:qualitative_comparison}
\end{figure}

\vspace{0.3em}
\noindent\textbf{Quantitative and Qualitative Analysis of GOP.}
We conduct ablation experiments to validate the effectiveness of the proposed GOP mechanism. As shown in Fig. \ref{fig:transferability_vis}, we first analyze the cosine similarity between the LoRA keys of different modules. Without GOP, the cosine similarities remain consistently high and are concentrated near 1, indicating strong directional overlap between LoRA updates. Such overlap implies that the watermark-carrying update is highly entangled with other adaptation directions, which can lead to interference during LoRA composition. In contrast, after applying GOP, the cosine similarities are sharply reduced and remain close to zero across the entire key set. This result quantitatively confirms that GOP effectively suppresses the first-order overlap between watermark updates and other LoRA-induced parameter directions.

The practical benefit of this reduced interference is further illustrated by the visual ablation in Fig. \ref{fig:qualitative_comparison}. Without GOP, the composed results exhibit noticeable degradation in both fidelity and style consistency. In the subject-driven example, the generated image suffers from structural distortion and noisy texture artifacts, while in the style-driven example, the result shows weakened stylistic coherence and degraded visual quality. By contrast, with GOP, the generated images preserve the intended subject structure, fine-grained details, and target stylistic characteristics much more faithfully. These observations are consistent with the quantitative similarity analysis, confirming that GOP improves modular compatibility by reducing parameter interference and thereby enables more stable watermark injection under LoRA composition.

\begin{figure}[t]
\centering
\includegraphics[width=\columnwidth]{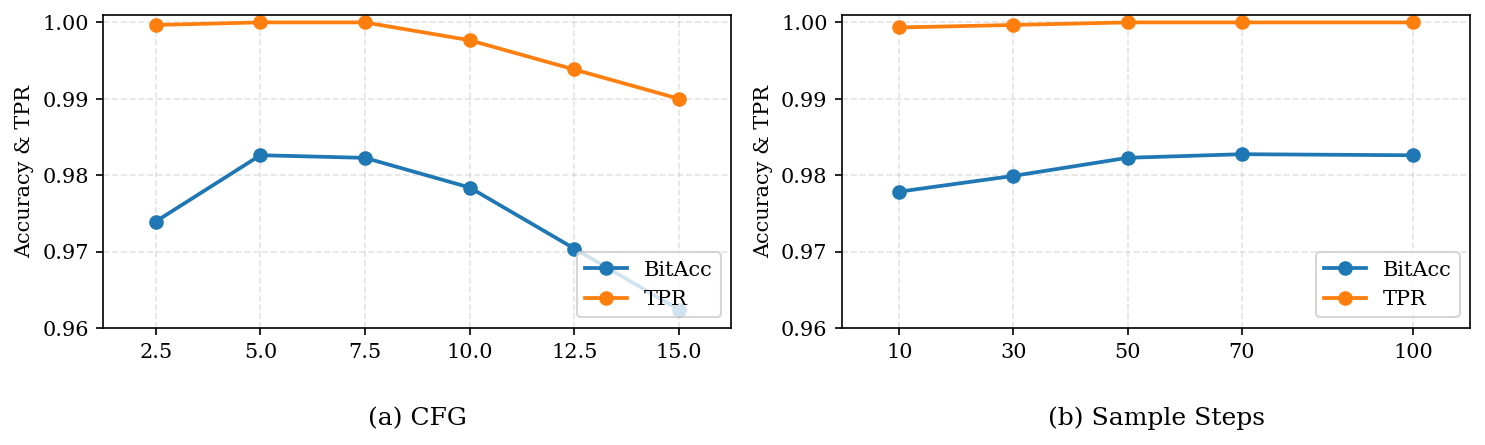}
\caption{Impact of inference hyperparameters on watermark extraction performance. (a) Effect of the classifier-free guidance (CFG) scale. (b) Effect of the number of sampling steps.}
\label{fig:cfg_steps}
\end{figure}

\vspace{0.3em}
\noindent\textbf{Sampling Steps and CFG Scale.}
We further analyze the effect of two inference hyperparameters, namely the classifier-free guidance (CFG) scale and the number of sampling steps, on watermark extraction performance. As shown in Fig.~\ref{fig:cfg_steps}(a), varying the CFG scale causes only minor fluctuations in Bit Accuracy, while the overall performance remains consistently high across the entire evaluated range. BitAcc reaches its best values at moderate CFG scales, approximately between 5 and 7.5, and decreases slightly at larger guidance strengths. Meanwhile, the TPR remains close to 1.00 throughout, indicating that LoRA-Key is highly robust to changes in CFG scale and can maintain reliable detection under different guidance settings.
Fig.~\ref{fig:cfg_steps}(b) shows a similar trend with respect to the number of sampling steps. BitAcc improves slightly as the number of sampling steps increases, and then quickly saturates after around 50 steps. The TPR also remains consistently high across all tested settings.

\section{Conclusion}
In this paper, we presented LoRA-Key, a modular and user-centric watermarking framework for protecting independently distributed LoRA assets in text-to-image diffusion models. By decoupling copyright protection from target LoRA training, LoRA-Key encapsulates ownership signals into a standalone Watermark LoRA that can be attached to diverse independently trained LoRA assets through training-free linear superposition. The framework integrates a latent watermark prior, GOP-based watermark optimization, and plug-and-play LoRA composition, enabling scalable deployment together with reliable ownership verification. Extensive experiments demonstrate that LoRA-Key preserves generation quality, semantic consistency, and style fidelity while maintaining robust watermark verification under image-level distortions, downstream fine-tuning, cross-architecture transfer, and multi-LoRA composition. These results show that LoRA-Key provides a practical and scalable solution for copyright protection in the emerging LoRA-centric diffusion ecosystem.




%
\bibliographystyle{IEEEtran}
\bibliography{sample-base}

\end{document}